\documentclass[12pt]{article}
\usepackage{a4,amsmath,amsxtra,amssymb}

\begin{document}

\baselineskip=1.5\baselineskip

\centerline{{{\LARGE Unfortunate Terminology}}}

\vskip12pt

\centerline{\large Jerzy Klemens Kowalczy\'nski}

\vskip16pt

\noindent
Mailing address: Institute of Physics, Polish Academy of Sciences,\\
Al. Lotnik\'ow 32/46, 02-668 Warsaw, Poland

\vskip12pt

\noindent
E-mail address: jkowal@ifpan.edu.pl

\vskip48pt

\noindent
{\bf Abstract.} The phrase ``negative squared rest mass'' can
sometimes be found in papers on neutrinos and frequently occurs
in the tachyonic literature. Consequently, some authors say
that ``the rest mass of tachyons is imaginary''. Besides, the
statement ``photons have zero rest mass'' is almost common.
In terms of relativity, however, the state of rest cannot be
reasonably defined for luxons and tachyons, and, therefore, today
it does not make sense to speak of rest mass of such objects.
It is shown here that the phrases ``negative squared
mass'', ``imaginary mass'', and ``photon's zero mass'' result
from applying bradyonic dynamical relativistic relations to
determine properties of luxons and tachyons; and that this
erroneous procedure results from an unfortunate interpretation
of kinematical relativistic relations. It is also shown that the
use of proper relativistic relations, i.e. luxonic or tachyonic
ones, gives a positive quantity having the squared mass
dimension. Thus we obtain a nonzero real quantity having the mass
dimension (called masslike quantity), which is positive for the
photon and may (by intuition -- should) be positive
for other luxons and for tachyons.

\newpage

From some papers on measurements of the
neutrino mass one can learn that ``the
squared rest mass of neutrinos is negative''.
In the tachyonic literature
it is frequently stated that ``the
squared rest mass of tachyons
is negative'', and, consequently, some
authors conclude that ``the rest mass of tachyons
is imaginary'', though in the case of neutrinos I have not met
such a heroic author.
Besides, the statement ``photons have zero rest mass'' is almost
common. These statements are said to be conclusions from
relativity but this is not true.

In relativity the term ``rest mass'' does not make sense in the
case of luxons and tachyons, since the state of rest can be
reasonably defined for these objects neither within standard
relativity nor in its consistent extensions. This is obvious in the
luxonic case since, e.g., the Lorentz transformation is singular
for speeds equal to~$ c $. If we were to assume that any
tachyon may be at rest, then three {\it independent} states of its
rest would have to exist since observers who see {\it our}
spacetime as that having three time dimensions and only one space
dimension would then have to be admitted.

As regards the phrases ``negative squared mass'', ``imaginary
mass'', and ``photon's zero mass'', we shall proceed step by
step.

Consider the world line $ x^{\mu}\left(\sigma\right) $ of a
pointlike object. Assume, for simplicity, that the object is free
in flat spacetime endowed with the Lorentzian coordinates (i.e.
$ x^{\mu}\left(\sigma\right) $ is straight), that $ \sigma $ is the
normalized affine parameter of $ x^{\mu}\left(\sigma\right) $, and
that the signature is, e.g., $ + + + - {}$. Note that in the metric
form expressions, $ ds^{2} = dx_{\mu}dx^{\mu},\ ds^{2} $ is only a
conventional symbol, and therefore it need not be the square of an
infinitesimal real quantity. In the case under consideration
$$
ds^{2} = dx^{2} + dy^{2} + dz^{2} - c^{2}dt^{2},\eqno (1)
$$
and for $ x^{\mu}\left(\sigma\right) $ we have
$$
ds^{2} = -k\left(d\sigma\right)^{2},\eqno (2)
$$
where $ d\sigma $ is indeed an infinitesimal real quantity, and
where the discrete dimensionless parameter $ k $ is as follows:

$k = 1$ in the bradyonic (timelike, subluminal) case,

$k = 0$ in the luxonic (null, luminal) case, and

$k = -1$ in the tachyonic (spacelike, superluminal) case.

\noindent (If the signature $ + - - - {}$
were chosen, then by Eq.~(2) we would have $ k = -1 $ in
the bradyonic case and $ k = 1 $ in the tachyonic case.) Dividing
Eqs. (1) and (2) by $
\left(d\sigma\right)^{2} $ we get
$$
-k = \left(u^{x}\right)^{2} + \left(u^{y}\right)^{2} +
\left(u^{z}\right)^{2} - \left(u^{t}\right)^{2},\eqno (3)
$$
where $ u^{\mu} := dx^{\mu}/d\sigma $ is a four-velocity vector.
The kinematical Eq.~(3) concerns every type of world lines --
timelike, null, and spacelike. The type is determined by~$ k $.

Multiplying Eq.~(3) by $ m^{2}c^{2} $, where $ m $ has
the mass dimension (we do {\it not yet\/} determine physical
meanings of~$ m $), we get the well-known special relativistic
formula for a four-momentum vector $ p^{\mu} $:
$$
-km^{2}c^{2} = \left(p^{x}\right)^{2} + \left(p^{y}\right)^{2} +
\left(p^{z}\right)^{2} - \left(p^{t}\right)^{2} \equiv {\bf p}^{2} -
c^{-2}E^{2},\eqno (4)
$$
where
$$
p^{\mu} := mcu^{\mu},\eqno (5)
$$
and where $ \left(p^{x}\right)^{2} + \left(p^{y}\right)^{2}
+ \left(p^{z}\right)^{2} \equiv {\bf p}^{2} $ and $
\left(p^{t}\right)^{2} \equiv c^{-2}E^{2} $. If we had $ m =
0$, then by definition (5) we would have no four-momentum, i.e.
no object on our world line (not speaking that
the multiplication of
equations by zero does not make sense). Thus
$$
m \neq 0.\eqno (6)
$$
If $ m $ were imaginary, then by definition (5) also the
four-momentum components $ p^{\mu} $ would be imaginary, which
would give us a new physics yet unknown.
If we had real $ m < 0$, then by definition
(5) we would have opposite senses of the four-vectors $ u^{\mu} $
and~$ p^{\mu} $. Such a situation is yet unknown and today seems
strange, though perhaps it will be considered in future. Anyway, we
are entitled to put real $ m > 0 $ for {\it every\/} type of the
objects under consideration (Ockham's principle!).

The unfortunate phrases have resulted from the fact that some
authors have not taken into account the existence of three values
of $ k $ (1, 0, $ -1 $) and have applied the bradyonic variants of
Eqs. (1)--(4) for luxons and tachyons.
The use of proper values of $ k $ allows to avoid the
difficulties.

In the bradyonic case, $ m $ is the rest mass of our object. In the
luxonic case the physical meaning of $ m $ is not determined in
general, though it is so for the photon for which $ m = c^{-2}E =
c^{-2}h\nu
> 0$. Anyway, the dynamical luxonic relation $ {\bf p}^{2}c^{2} =
E^{2} $ does {\it not\/}
result from the condition $ m = 0$, which is false (inequality~(6)),
but it does result from the condition $ k = $ 0, i.e. it is
determined at the {\it kinematical\/} level of Eqs. (1)--(3).
In the tachyonic case we have yet no operational definition of $ m
$ (for lack of rest), and therefore the term ``masslike quantity''
has been proposed.
(The terms ``pseudomass'' or ``quasimass'' are shorter
but semantically inferior.)

Additional remarks and amusing details can be found in Section 6
of my prior paper [1].

\vskip20pt

\noindent
[1] J. K. Kowalczy\'nski, Acta Physica Slovaca {\bf 50} (2000) 381,
or hep-ph/9911441.

\end{document}